\documentclass[a4paper]{article}

\usepackage{fullpage}
\usepackage{graphicx}
\usepackage{amssymb}
\usepackage{amsmath}
\usepackage{subfigure}
\usepackage{hyperref}
\usepackage{empheq}   
\usepackage{color}

\newcommand*{\affaddr}[1]{#1} 
\newcommand*{\affmark}[1][*]{\textsuperscript{#1}}

\definecolor{orange}{rgb}{0.93, 0.53, 0.18}

\begin{document}



\title{\textit{Thermoelectric Effects in Nanowire-Based MOSFETs}}

\author{
Riccardo Bosisio\affmark[1], Genevi\`eve Fleury\affmark[1], Cosimo Gorini\affmark[1,2] and Jean-Louis Pichard\affmark[1]$^{\ast}$\thanks{$^\ast$Corresponding author. Email: jean-louis.pichard@cea.fr}\\
\affaddr{\affmark[1] \small{SPEC, CEA, CNRS, Universit\'e Paris-Saclay, CEA-Saclay, 91191 Gif-sur-Yvette, France}}\\
\affaddr{\affmark[2] \small{Institut f\"{u}r Theoretische Physik, Universit\"{a}t Regensburg, 93040 Regensburg, Germany}}
}

\maketitle

\begin{abstract}

We review a series of works describing thermoelectric effects in gated disordered nanowires (field effect transistor device configuration). 
After considering the elastic coherent regime characterizing sub-Kelvin temperatures, we study the inelastic activated regime occurring at 
higher temperatures, where electronic transport is dominated by phonon-assisted hops between localised states (Mott variable range hopping). 
The thermoelectric effects are studied as a function of the location of the Fermi level inside the nanowire conduction band, notably around 
its edges where they become very large. We underline the interest of using electron-phonon coupling around the band edges of large arrays of 
parallel nanowires for energy harvesting and hot spot cooling at small scales. Multiterminal thermoelectric transport and ratchet effects are 
eventually considered in the activated regime.  

\end{abstract}

\section{Introduction}

 Let us consider a nanowire (NW) connecting two electron reservoirs. If one imposes a temperature difference $\delta T$ 
between the reservoirs, this induces an electrical current $I_e$ which can be suppressed by a voltage difference $-\delta V$. 
The ratio $S=-(\delta V/\delta T)_{I_e=0}$ defines the NW Seebeck coefficient (or thermopower). If one imposes a voltage 
difference $\delta V$ when $\delta T=0$, this induces  electrical and heat currents $I_e$ and $I_Q$. The ratio $\Pi=I_Q/I_e$ 
defines the NW Peltier coefficient. In the linear response regime, the Peltier and Seebeck coefficients are related via the 
Kelvin-Onsager relation $\Pi=ST$.     

Either a temperature gradient across a NW can produce electricity (Seebeck effect), or an electric current through the same 
NW can create a temperature difference between its two sides (Peltier effect). These thermoelectric effects (TEs) can be used 
either for harvesting electrical energy from wasted heat or for cooling things. Today, the batteries of our cell phones and laptops 
need to be charged too often. Tomorrow, the Seebeck effect could allow us to exploit the wasted heat to produce a part of the 
electrical energy necessary for many devices used for the internet of things. Another important issue is cooling, notably the hot 
spots in microprocessors. The last decades have been characterized by an exponential growth of the on-chip power densities. Values 
of the order of $100 W/cm^2$ have become common~\cite{Pop et al}. More than our ability to reduce their sizes, the limitation of 
the performances of microprocessors comes from the difficulty of managing heat in ever-smaller integrated circuits. Improving 
Peltier cooling and heat management from the nanoscale (e.g. molecules) to the microscale (e.g. quantum dots and nanowire arrays) 
is thus of paramount importance to boost microprocessors' performance. 

In a typical two-terminal configuration in which a device is coupled to two electronic reservoirs held at different temperatures, 
the ratio $\eta$ of the output power over the heat extracted from the hot reservoir measures the efficiency of the heat-to-work thermoelectric 
conversion. It cannot exceed the Carnot efficiency $\eta_C=1-T_C/T_H$, where $T_C$ ($T_H$) is the temperature of the cold (hot) 
reservoir. The figure of merit $ZT$ gives the maximal efficiency $\eta_{max}$ in terms of the Carnot limit~\cite{ioffebook,goldsmid}.   
\begin{equation}
ZT=\frac{GS^2}{K^{e}+K^{ph}} T; \ \ \ \ \ \ \eta_{max}=\eta_C\frac{\sqrt{ZT+1}-1}{\sqrt{ZT+1}+1},
\end{equation} 
where $G$ is the electrical conductance, while $K^{e}$ and $K^{ph}$ are respectively the electronic and phononic parts of the thermal 
conductance. The larger $ZT$, the better the efficiency. A high efficiency is however mainly useful if coupled with good electrical 
output power, measured by the power factor $Q=GS^2$. Maximising both $ZT$ and $Q$ is the central challenge of (linear response) 
thermoelectricity. This is not easy since $G$, $K_e$, $K_{ph}$ and $S$ are not independent.

\begin{figure}
  \centering
  \includegraphics[keepaspectratio,width=\columnwidth]{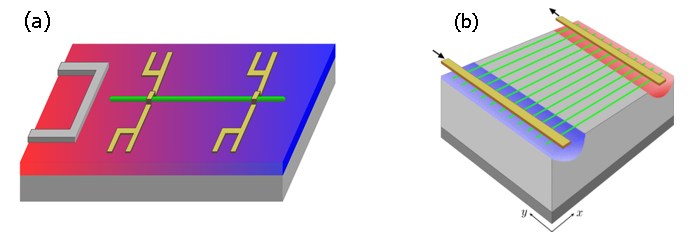}
  \caption{\label{FIG1} 
NW-based MOSFETs: ({\bf a}) A single NW (green) is deposited on an insulating substrate (red and blue). The source and the drain are made of two 
metallic electrodes (yellow), while Joule heating from an extra electrode (left side) can induce a temperature difference between the 
NW extremities. Varying the voltage $V_g$ applied upon the back gate (grey), one can shift the NW conduction band and probe thermoelectric 
transport in the bulk of the band, around its edges or even outside the band. This setup has been used in Ref.~\cite{Kim} for 
measuring the thermopower ${\cal S}$ of individual Si and Ge/Si NWs as a function of $V_g$ at room temperature. ({\bf b}) 
Array of parallel NWs deposited on a substrate with a back gate. The blue and red spots illustrate local cooling and heating effects in 
the activated regime, discussed in Sec.\ref{section_HotSpot}.} 
\end{figure}

 The interest of NWs for thermoelectric conversion was pointed out in Ref.~\cite{Hochbaum}. Taking arrays of parallel doped Si NWs 
of $50$ nm in diameter yields $ZT=0.6$ at room temperature, a much larger value than in bulk silicon. This was attributed to a $100$-fold 
reduction in thermal conductivity, assuming than $S$ and $G$ keep the same values than in doped bulk Si. Using standard Si-based semiconductor 
technology for thermoelectric conversion looks very interesting: In contrast to used thermoelectric materials, Si is cheap and non toxic, and 
NW-based one dimensional (1D) electronics is a well developed technology. Moreover, one can use metallic gates for tuning the NW electron 
density, in the field effect transistor (FET) device configuration. A detailed experimental study of electron tunneling and interferences in 1D 
Si-doped Ga-As MOSFETs can be found in Ref.~\cite{Poirier}, where the electrical conductance $G$ of $1 \mu m$ long NWs at $35 mK$ is given as a 
function of the gate voltage $V_g$. One can make~\cite{Eymery} arrays of vertical NW-based FET, each of them having a uniform wrap-around gate. 
To grow millions of thin NWs per $cm^2$ is possible. This gave us the motivation to study the TEs in 1D MOSFETs, from cryogenic temperatures where 
electron transport remains coherent towards higher temperatures where transport becomes activated, as a function of the location of the Fermi 
potential $E_F$ inside the NW conduction band. The considered setups are sketched in Fig.~\ref{FIG1}. 

There are many highly cited works describing the thermoelectric performance of NWs made of different materials: rough silicon~\cite{Hochbaum,Boukai}, 
bismuth~\cite{Sun}, bismuth telluride, III-V semiconductors (InP, InAs, and GaAs)~\cite{Mingo}, wide band gap semiconductors 
(ZnO and GaN)~\cite{Lee}. These studies have essentially been done around room temperature, and are mainly focused on the study of 
the phononic contribution to the thermal conductance~\cite{Hochbaum,Boukai} or of the effect of channel openings occuring when one 
varies the widths of superlattice quasi-1D NWs~\cite{Sun,Lin}. Thermoelectric transport is often described using 1D Boltzmann 
equations~\cite{Mingo,Sun,Lee}, and the role of localised impurity states (important in weakly doped semiconductors) as well as the 
Anderson localisation of the states (important in the 1D-limit) is not taken into account in these studies.     

In this short review, we describe the effect of 1D Anderson localisation upon 1D thermoelectric transport, as one varies the location of the 
Fermi potential $E_F$ inside the NW conduction band. For this purpose, we use a purely 1D model (1D Anderson model) where the energy dependence 
of the localisation length and of the density of states is analytically known in the weak disorder limit. Though it does not allow us to 
describe the effect of channel openings occuring as one varies the NW width as in Refs.~\cite{Sun,Lin}, these studies describe thermoelectric 
transport in NWs where the one body states are localised. A low temperature elastic regime is considered where the conductance and the thermopower 
are respectively obtained from the Landauer and Mott formulas, followed by the study of an inelastic regime occuring at higher temperatures and 
characterised by phonon-activated hopping between localised states (Mott variable range hopping). In this activated regime, thermoelectric transport 
is not described using semi-classical Boltzmann equations, but from the numerical solution of the random resistor network model introduced by Miller 
and Abrahams for describing inelastic activated transport. 

 \section{Elastic thermoelectric transport}
 
To model a gated NW, we have considered in Ref.~\cite{bosisio1} a chain of $N$ sites coupled to two electronic reservoirs $L$ (left) and $R$ (right), 
in equilibrium at temperature $T_L=T+\delta T$ [$T_R=T$] and chemical potential $\mu_L=E_F+\delta\mu$ [$\mu_R=E_F$]. The Hamiltonian of the 
chain reads
\begin{equation}
\label{eq_modelAnderson1D}
\mathcal{H}=-t\sum_{i=1}^{N-1}\left(c_i^{\dagger}c_{i+1}+\text{h.c.}\right)+\sum_{i=1}^{N}(\epsilon_i+V_g) c_i^{\dagger}c_i\,,
\end{equation} 
where $c^{\dagger}_i$ and $c_i$ are the creation and annihilation operators of one electron on site $i$ and $t$ is the hopping energy. 
The lattice spacing $a=1$, the $\epsilon_i$ are (uncorrelated) random numbers uniformly distributed in the interval $[-W/2,W/2]$. 
$\sum_i V_g c_i^{\dagger}c_i$ describes the effect of an external gate. Varying $V_g$, one can probe thermoelectric transport either 
in the bulk of the NW conduction band, around its edges or even outside the band. 

\subsection{Typical thermopower}
\label{elastic-thermopower}

For the Hamiltonian \eqref{eq_modelAnderson1D}, the localisation length $\xi(E)$ and the density of states (DOS) $\nu(E)$ 
per site are analytically known~\cite{Derrida-Gardner} in the weak disorder limit $W \leq t$. Within the band  
($|E-V_g|\lesssim 1.5t$), $\xi(E)^{-1}$ can be expanded in integer powers of $W$ while $\nu(E)$ remains well described 
by the DOS of the clean chain ($W=0$). This gives the bulk expressions 
\begin{equation}
\label{Bulk}
\xi_b(E)\approx \frac{24}{W^2}\left(4t^2-|E-V_g|^2\right), \ \ \ \ \ \nu_b(E)\approx \frac{1}{2\pi t\sqrt{1-(|E-V_g|/2t)^2}}.  
\end{equation}
When $E-V_g$ approaches the band edges $\pm 2t$, these expressions lead to divergences of $\nu$ and $\xi^{-1}$. 
As shown by Derrida and Gardner, these divergences are spurious and the correct expressions near the edges become
\begin{eqnarray}
\xi_e(E) &=& 2\left(\frac{12t^2}{W^2}\right)^{1/3}\frac{\mathcal{I}_{-1}(X)}{\mathcal{I}_{1}(X)} \label{Edge1} \\
\nu_e(E) &=& \sqrt{\frac{2}{\pi}}\left(\frac{12}{tW^2}\right)^{1/3}\frac{\mathcal{I}_1(X)}{[\mathcal{I}_{-1}(X)]^2} \label{Edge2}
\end{eqnarray}
where 
\begin{equation}
X=(|E-V_g|-2t)t^{1/3}(\frac{12}{W^2})^{2/3}, \ \ \ \ \mathcal{I}_n(X)=\int_0^{\infty} y^{n/2}\,e^{-\frac{1}{6}y^3+2Xy}\,dy\,.
\label{eq_scaling-variable}
\end{equation}

\begin{figure}
  \centering
  \includegraphics[keepaspectratio,width=\columnwidth]{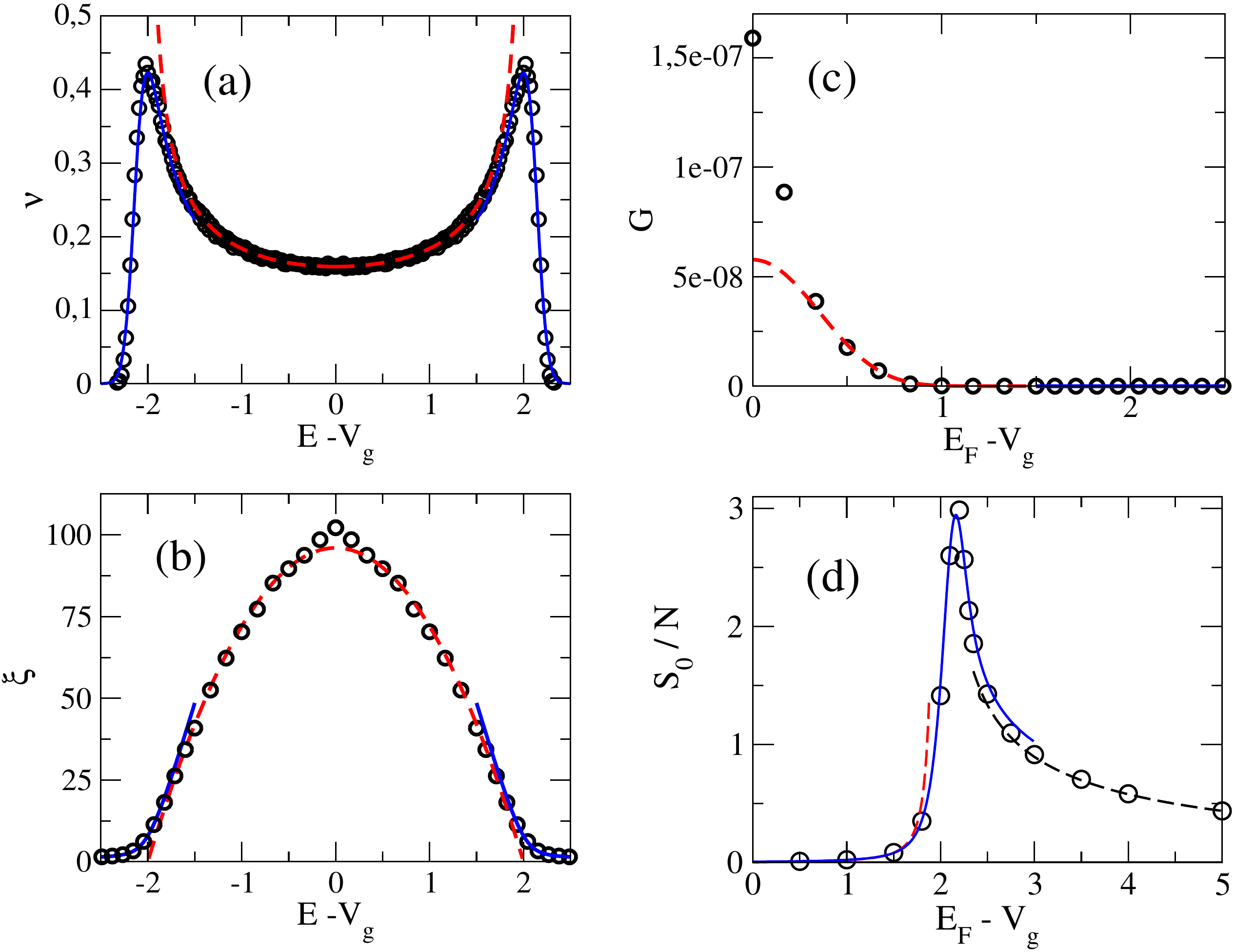}
  \caption{1D Anderson model with $W=t=1$: ({\bf a}) Density of states per site $\nu(E)$, ({\bf b}) localisation length 
$\xi(E)$, ({\bf c}) typical conductance $G$ (in units of $2e^2/h$) and ({\bf d}) typical thermopower $S_0$ (in units of $(\pi^2k_B)/(3e)\,k_BT$). In all panels, the red dashed line and the blue continuous line give the weak disorder behaviours in the bulk and near the edges
(Eqs.\,(\ref{Bulk}-\ref{Edge2}) and Eqs.\,(\ref{eq_S0bulk}-\ref{eq_S0edge})), while the black dashed line in (d) corresponds to Eq.\,\eqref{eq_S0TB}. Circles are numerical results obtained for $N=1600$ ((a) and (b)) and $N=800$ ((c) and (d)).}
\label{FIG2}
\end{figure}

The transmission coefficient $\mathcal{T}(E)$ of the disordered chain behaves typically as $\exp -2N/\xi(E)$. In the low temperature limit $T \rightarrow 0$, the electrical conductance $G\approx\frac{2e^2}{h}\mathcal{T}(E_F)$ while the thermopower $S$ is given by the Cutler-Mott formula, 
\begin{equation}
\label{eq_SeebeckMott}
S=\frac{\pi^2 k_B^2 T}{3 |e| t} \mathcal{S}~~\mathrm{with}~~\,\mathcal{S}\approx-t\left.\frac{\mathrm{d}\ln\mathcal{T}}{\mathrm{d}E}\right|_{E_F}\,.
\end{equation} 
From the weak disorder expansions of $\xi(E)$, one can deduce the typical dimensionless thermopower $\mathcal{S}_0$ (thermopower in units of $(\pi^2 k_B^2 T)/(3 |e| t)$). This gives respectively in the bulk of the band (superscript $b$) and at its edges (superscript $e$):
\begin{equation}
\label{eq_S0bulk}
\mathcal{S}_0^b=N \frac{(E_F-V_g)\,W^2}{96t^3[1-((E_F-V_g)/2t)^2]^2},
\end{equation} 
\begin{equation}
\label{eq_S0edge}
\mathcal{S}_0^e=2N \left(\frac{12t^2}{W^2}\right)^{1/3}\left\{ \frac{\mathcal{I}_{3}(X)}{\mathcal{I}_{-1}(X)}
-\left[\frac{\mathcal{I}_{1}(X)}{\mathcal{I}_{-1}(X)}\right]^2\right\},
\end{equation} 
where $X=X(E=E_F)$. Outside the band, one estimates the typical thermopower by assuming that the system behaves as a clean tunnel barrier (superscript $TB$). One obtains
\begin{equation}
\label{eq_S0TB}
\frac{\mathcal{S}_0^{TB}}{N} \underset{N\to\infty}{\approx} -\frac{1}{N}\frac{2t}{\Gamma(E_F)}
\left.\frac{\mathrm{d}\Gamma}{\mathrm{d}E}\right|_{E_F}\mp\frac{1}{\sqrt{\left(\frac{E_F-V_g}{2t}\right)^2-1}}
\end{equation} 
with a $+$ sign when $E_F\leq V_g-2t$ and a $-$ sign when $E_F\geq V_g+2t$. Here $\Gamma(E)=i[\Sigma(E)-\Sigma^\dagger(E)]$ where $\Sigma(E)$ is the self-energy of the (identical) left and right leads, evaluated at the sites located at the chain extremities to which the leads are attached.
In Fig.~\ref{FIG2}, one can see that the analytical weak disorder expressions of the DOS per site $\nu(E)$, of the localisation length 
$\xi(E)$, of the electrical conductance $G(E_F)$ and of the typical thermopower $S_0(E_F)$ (in units of $(\pi^2 k_B^2 T)/(3 |e| t)$) describe accurately numerical results 
(for more details, see Ref.~\cite{bosisio1}), even if they are computed for a relatively large disorder ($W=t$). 

\subsection{Mesoscopic Fluctuations}
\label{section_distrib}

In the elastic localized regime, the sample-to-sample fluctuations of the dimensionless thermopower $\mathcal{S}$ around its typical values $\mathcal{S}_0^b$ 
or $\mathcal{S}_0^e$ turn out to be large. If $E_F=V_g$, $\mathcal{S}_0=0$  due to particle-hole symmetry but the mesoscopic fluctuations allow for a 
large $\mathcal{S}$ anyway. Assuming Poisson statistics for the energy levels, Van Langen \textit{et al} showed in Ref.~\cite{VanLangen1998} that 
the thermopower distribution is a Lorentzian when $N\gg\xi$,
\begin{equation}
\label{eq_distrib_lor}
P(\mathcal{S})=\frac{1}{\pi}\frac{\Lambda}{\Lambda^2+(\mathcal{S}-\mathcal{S}_0)^2}\, .
\end{equation} 
The width $\Lambda=2\pi t/\Delta_F$ is given by the mean level spacing $\Delta_F=1/(N \nu(E_F))$ at the Fermi energy $E_F$. 
Van Langen \textit{et al} assumed $\mathcal{S}_0=0$, an assumption which is only correct at the band centre. We have calculated 
$P(\mathcal{S})$ using recursive Green function method for different values of $V_g$ and have numerically checked~\cite{bosisio1} that 
Eq.~\eqref{eq_distrib_lor} describes also $P(\mathcal{S})$ if one takes for $\mathcal{S}_0$ the value given by Eqs.~\eqref{eq_S0bulk} 
or ~\eqref{eq_S0edge} instead of $\mathcal{S}_0=0$. As one crosses the band edges, we have numerically observed a sharp crossover 
towards a Gaussian distribution
\begin{equation}
\label{eq_distrib_gauss}
P(\mathcal{S})=\frac{1}{\sqrt{2\pi}\lambda}\exp\left[-\frac{(\mathcal{S}-\mathcal{S}_0)^2}{2\lambda^2}\right]\,,
\end{equation} 
where the typical value $\mathcal{S}_0$ is given by Eq.~\eqref{eq_S0TB} and the width $\lambda$ increases linearly with 
$\sqrt{N}$ and $W$. In Ref.~\cite{bosisio1}, one can find numerical results which are perfectly described by 
the above analytical expressions when $W=t$.

\section{Inelastic thermoelectric transport}
\label{section_inelastic}

\begin{figure}
  \centering
 \includegraphics[keepaspectratio, width=0.9\columnwidth]{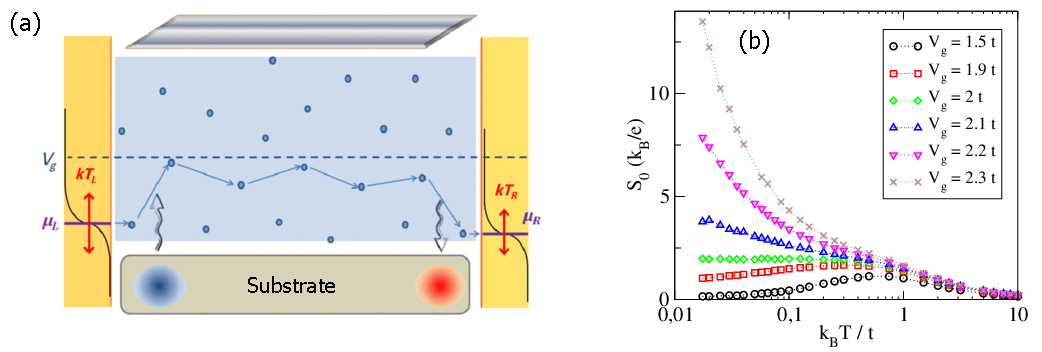}
\caption{({\bf a}) The electronic localised states (blue dots) are randomly located along the chain within an energy band of width $4t+W$ 
(shaded light blue region). In the left and right reservoirs, the electrons are at equilibrium with Fermi-Dirac distributions of temperatures 
$T_L=T+\delta T$, $T_R=T$ and electrochemical potentials $\mu_L=E_F+\delta\mu$, $\mu_R=E_F$. Varying the voltage $V_g$ applied to the gate 
(top grey region) shifts the NW conduction band, making possible to probe electron transport inside the band, around its edges or outside the band. 
The electrons can hop between states of different energies by absorbing or emitting a phonon. For a NW deposited on a substrate, the phonons are 
provided by the substrate and characterized by a Bose-Einstein distribution of temperature $T_S$ ($T_S=T$ throughout the paper, except in 
Sec.\ref{section_ratchet}). Electrons injected near the lower band edge mainly find available states to jump to at higher energies near the source 
(left reservoir). This implies that the phonons are mainly absorbed from the substrate near the source (blue region), and emitted back to the substrate 
near the drain (red region), as indicated by the two arrows. ({\bf b}) Typical thermopower $S_0$ (in units of $k_B/e$) as a function of $k_BT/t$, 
calculated for a chain of length $L=200$ ($E_F=0$ and $W=t$) for increasing values of $V_g$.}
\label{FIG3}   
\end{figure}

 When one increases the temperature $T$, electron transport becomes mainly inelastic and activated. The inelastic effects can be due to 
electron-electron, electron-photon and electron-phonon interactions. Electron-electron interactions in a many-electron system with localised 
single-particle states can induce a metal-to-insulator transition above a certain critical temperature~\cite{Basko}. These interactions can also 
induce the Coulomb-gap behaviour~\cite{Efros-Shklovskii} observed in $\delta$-doped $GaAs/Al_xGa_{1-x}As$ 2D heterostructures at low 
temperature~\cite{Khondaker}, providing evidence of possible phononless hopping. Electron-photon interactions are responsible for the photovoltaic 
effects. In Ref.~\cite{bosisio2}, we have studied the Variable Range Hopping (VRH) regime introduced by Mott~\cite{Mott1979} where electron-phonon 
coupling dominates. Fig.\ref{FIG3} (a) illustrates how electrons propagate through the NW in the Mott VRH regime.

\subsection{Variable Range Hopping}
\label{VRH}
 
In the VRH regime, the electrons propagate by hopping from one localised state of energy $E_i$ to another, of higher energy $E_j>E_i$ by absorbing a phonon or of lower energy $E_j<E_i$ by emitting a phonon. Let us summarize Mott's original 
argument~\cite{Mott1979}. The electron transfer from a state $i$ to another state $j$ separated by a distance $L_{ij}$ in space and $\Delta_{ij}$ 
in energy results from a competition between the probability $\propto \exp -(L_{ij}/\xi)$ to tunnel over a length $L_{ij}$ and the probability 
$\propto \exp -\Delta_{ij} /(k_BT)$ to change the electron energy by an amount $\Delta_{ij}=1/(\nu L_{ij})$ , where $\nu$ is the DOS per site. 
These estimates neglect the energy dependence of $\xi$ and $\nu$ around $E_F$. In 1D, the optimal hopping length is given by the \textit{Mott length} 
$L_M\simeq (\xi/2 \nu k_BT)^{1/2}$, if the localisation lengths $\xi$ and DOS per unit length $\nu$ do not vary within the Mott 
energy window $\Delta_M=1/(\nu L_M)=k_B\sqrt{TT_M}$ around $E_F$. $L_M$ decreases as the temperature increases. One defines 
the activation temperature $k_BT_x\simeq \xi/(2\nu L^{2})$ at which $L_M\simeq L$ and the Mott temperature $k_BT_M\simeq 2/(\nu \xi)$ at which 
$L_M \simeq \xi$. The inelastic VRH regime corresponds to $T_x<T<T_M$ where the electrical conductance 
\begin{equation}
G \propto \exp -(2L_M/\xi) \propto \exp -(\Delta_M/k_BT).  
\end{equation}
Below $T_x$ elastic tunneling dominates, while $L_M < \xi$ above $T_M$ and transport becomes simply activated. In 1D, the crossover 
from VRH to simply activated transport takes even place ~\cite{Kurkijarvi1973,Raikh1989} at a temperature $T_a$ lower than $T_M$.  
The reason is the presence of highly resistive regions in energy-position space, where 1D electrons cannot find empty states at distances 
$\sim\Delta_M, L_M$.

\subsection{Random Resistor Network with energy-dependent localisation length and density of states}
\label{RRN}

If the variation of $\xi(E)$ and $\nu(E)$ as a function of the energy $E$ is not negligible within the characteristic scale $\Delta_M$, we need to go 
beyond this simple argument, notably around the 1D band edges. We use a simplified model where the $E_i$ are $N$ uncorrelated variables of probability 
given by the DOS $\nu(E)$ of the 1D Anderson model for a chain of length $L=Na$ ($a=1$), while the $N$ localisation lengths $\xi(E_i)$ are given by the 
typical values of this model (Eqs.~\eqref{Bulk} and~\eqref{Edge1}). The $N$ positions $x_i$ of the localized states are taken at random in the interval $[0,L]$. As in 
Refs.~\cite{Jiang2012,Jiang2013}, we solve the corresponding Miller-Abrahams random resistor network~\cite{Miller1960} (RRN) made of all possible links 
connecting the $N$ nodes given by the $N$ localised states. Each pair of nodes $i, j$ is connected by an effective resistor, which depends on the 
transition rates $\Gamma_{ij}, \Gamma_{ji}$ induced by local electron-phonon interactions. For a pair of localised states $i$ and $j$ of energies $E_i$ 
and $E_j$, Fermi golden rule~\cite{Jiang2013} gives:
\begin{equation}
\label{eq:gamma_ij}
\Gamma_{ij}=\gamma_{ij}\,f_i\,(1-f_j)\,\left[N_{ij}+\theta(E_i-E_j)\right]\,,
\end{equation}
where $f_i$ is the occupation number of state $i$ and $N_{ij}=[\text{exp}\{|E_j-E_i|/k_BT\}-1]^{-1}$ 
is the phonon Bose distribution at energy $|E_j-E_i|$.  The Heaviside function accounts for the difference between phonon absorption 
and emission~\cite{Ambegaokar1971}. $\gamma_{ij}$ is the hopping probability $i \to j$ due to the absorption/emission of a phonon when 
$i$ is occupied and $j$ is empty. It is given by
\begin{equation}
\frac{\gamma_{ij}\left( 1/\xi_i - 1/\xi_j\right)^{2}}{\gamma_{ep}}=\frac{\exp\{-2x_{ij}/\xi_j\}}{\xi_i^2}+ 
\frac{\exp\{-2x_{ij}/\xi_i\}}{\xi_j^2} -\frac{2\exp\{-x_{ij} (1/\xi_i + 1/\xi_j)\}}{\xi_i \xi_j} 
\label{eq:gamma_ij_app_4}
\end{equation}
where $x_{ij}=|x_i-x_j|$ and $\gamma_{ep}$ depends on the electron-phonon coupling strength and of the phonon density of states. 
If the energy dependence of $\xi$ and $\nu$ can be neglected within $\Delta_M$, one recovers the usual limit 
$\gamma_{ij}\simeq \gamma_{ep}\,\text{exp}(-2x_{ij}/\xi)$. 
\\ 
\indent 
The direct transition rates between each state $i$ and the contacts $\alpha$ (source $\alpha=L$ and drain $\alpha=R$) are 
assumed to be dominated by elastic tunneling (see Refs.~\cite{Jiang2012,Jiang2013}) and read 
\begin{equation}
\Gamma_{i\alpha} = \gamma_{e,\alpha}\,\text{exp}(-2x_{i\alpha}/\xi_i)\,f_i\,\left[1-f_{\alpha}(E_i)\right]. 
\label{elastic-coupling}
\end{equation}
$f_{\alpha}(E)=[\text{exp}\{(E-\mu_\alpha)/k_BT\}+1]^{-1}$ is the contact $\alpha$'s Fermi-Dirac distribution, $x_{i\alpha}$ denotes the 
distance of the state $i$ from $\alpha$, and $\gamma_{e,\alpha}$ is a rate quantifying the coupling between the localized states and the 
contact $\alpha$. The electric currents flowing between each pair of states and between states and contacts read
\begin{subequations}
\label{eq:currents}
\begin{align}
I_{ij} &= e\,(\Gamma_{ij}-\Gamma_{ji})\label{eq:currents_1},\\
I_{i\alpha} &= e\,(\Gamma_{i\alpha}-\Gamma_{\alpha i}),\,\qquad\,\alpha=L,R\label{eq:currents_2}.
\end{align}
\end{subequations}
$e<0$ is the electron charge. Hereafter, we will take $\gamma_{ep}=t/\hbar$ and symmetric couplings $\gamma_{e,L}=\gamma_{e,R}=t/\hbar$.

 For solving the RRN, we consider it at equilibrium with a temperature $T$ and a chemical potential $\mu=E_F$ everywhere.  
A small electric current $I_e$ can be driven by adding to the left contact (the source) a small increase $\delta \mu$ of its chemical 
potential (Peltier configuration). If $\delta \mu$ is sufficiently small, one has $I_e \propto \delta \mu$ (linear response). 
At equilibrium ($\delta \mu=0$), the $N$ occupation numbers $f_i$ are given by Fermi-Dirac distributions $f_i^0=(\exp[(E_i-E_F)/k_BT] +1)^{-1}$. 
When $\mu_L \rightarrow E_F+\delta \mu$, $f_i \rightarrow f_i^0+\delta f_i$. For having the currents $I_{ij}$ and $I_{i\alpha}$, we only need to 
calculate the $N$ changes $\delta f_i$ induced by $\delta \mu \neq 0$. Imposing current conservation at each node $i$ of the network 
($ \sum_{j} I_{ij}+\sum_{\alpha}I_{i\alpha}=0$) and neglecting terms $\propto \delta f_i. \delta f_j$ (linear response), one obtained  
$N$ coupled linear equations. Solving numerically this set of equations gives the $N$ changes $\delta f_i$ and hence all the currents 
$I_{ij}$ and $I_{i\alpha}$. From this, we can calculate the total charge $I^e_L=-\sum_iI_{iL}$ and heat $I^Q_{L(R)}=\sum_i(E_i-\mu_{L(R)})/e)I_{iL(R)}$ 
currents, and hence the electrical conductance $G$, the Peltier coefficient $\Pi$ and the Seebeck coefficient $S$ in the VRH regime.
\begin{equation}
 G = \frac{I^e_L}{\delta \mu/e},\ \ \ \Pi = \frac{I^Q_L}{I^e_L},\ \ \ S =\frac{1}{T}\,\frac{I^Q_L}{I^e_L}. 
\label{eq:GPIS}
\end{equation}
In the last equation, the Kelvin-Onsager relation $\Pi=ST$ has been used for obtaining the thermopower $S$ from the Peltier coefficient $\Pi$. 

\subsection{Activated thermoelectric transport in arrays of parallel NWs}
\label{RRN_result}

 Using the 1D weak-disorder expressions (Eqs.\,(\ref{Bulk})-(\ref{Edge2})) for $\nu(E)$ and $\xi(E)$, we have studied activated transport 
through $N$ localised states of energy $E_i$ and localisation length $\xi(E_i)$. The states were assumed to be randomly located 
along a chain of length $L=Na$, and the energies $E_i$ were taken at random with a probability $\nu(E)$ inside an energy band 
$[-2 \epsilon, 2 \epsilon]$ where $\epsilon=t+W/4$. The corresponding thermopower distributions $P(S)$ are given and discussed in 
Ref.~\cite{bosisio2}. We reproduce in Fig.~\ref{FIG3} ({\bf b}) the curves giving the typical thermopower $S_0$ (in units of $k_B/e$) 
as a function of $k_BT/t$ for increasing values of $V_g$. Taking $E_F=0$, $S_0$ has been calculated for a chain of length $L=200$ with 
$W=t$. $V_g=0$ corresponds to the band centre and $V_g/t=\pm 2.5$ to the band edges. When $k_BT < t$, $S_0$ remains small within the band 
($V_g \approx 1.5 t$), but becomes much larger around its edges ($V_g=2.3 t$). At higher temperatures, $S_0$ decreases and becomes independent 
of $V_g$. If activated transport at the band edges give rise to large thermopowers, it is also characterized by small electrical conductances, 
which defavor large values for the power factors $Q$. This led us to consider in Ref.~\cite{bosisio3} arrays of parallel nanowires in the field 
effect transistor device configuration. In such arrays, the conductances add while the thermopower fluctuations self-average. The maximal output power $P_{max}=Q(\delta T)^2/4$ is found to be maximal near the band edges, while the electronic figure of merit (obtained without including the phononic contribution $K^{ph}$ to the thermal conductance) keeps a large value $Z_eT \approx 3$.
As estimated in Ref.~\cite{bosisio3}, $P_{max}$ can be of the order of $P_{max} \approx 20 {\tilde \gamma_e} \mu W$ for $M=10^5$ 
parallel silicon NWs with $\delta T \approx 10 K$, $T \approx 100 K$, and $t/k_B\approx 150$\,K. The larger is $M$ or $\delta T$, the larger is $P_{max}$.  
Estimates of the constant ${\tilde \gamma_e}=\gamma_e\hbar/t$ give values  $\approx 0.01 -1$. 
If one takes into account $K^{ph}$, we expect that $ZT \approx  Z_eT/(1+2/{\tilde \gamma_e})$ for Silicon suspended NWs, while $ZT \approx  
Z_eT/(1+20/{\tilde \gamma_e})$ for Silicon NWs deposited on a Silicon Dioxide substrate. 

\section{Using electron-phonon coupling for managing heat}
\label{section_HotSpot}
\begin{figure}
  \centering
  \includegraphics[keepaspectratio,width=\columnwidth]{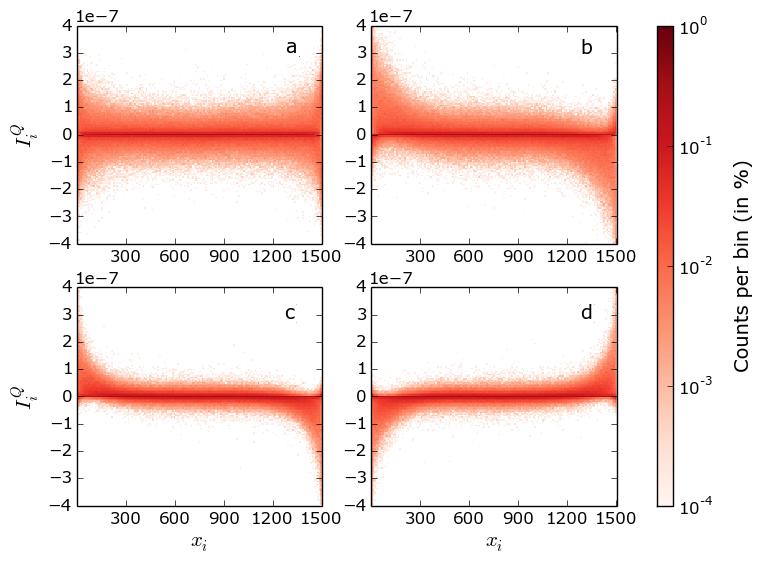}
  \caption{\label{FIG4} 
2D histograms giving the distribution of local heat currents $I^Q_i$ (Eq.~\eqref{local-current}) as a function of the position $x_i$ along the NW, 
for four values of the gate potential: (a) $V_g=0$, (b) $V_g=t$, (c) $V_g=2.25t$ and (d) $V_g=-2.25t$. Parameters: $L=1500$, $W=t$, $k_BT=0.5t$, 
$\delta\mu=10^{-5}t$, statistics over $500$ NWs.  
} 
\end{figure}

The phonons have no charge and cannot be manipulated with bias and gate voltages, in contrast to electrons. This makes difficult 
to manage heat over small scales, unless we take advantage of the electron-phonon coupling for transfering heat from the phonons 
towards the electrons. Let us show how this can be done using phonon-activated transport near the edges of a NW conduction band. 
We take a NW deposited on a substrate with a back gate, and assume that the transport of NW-electrons is activated mainly because 
of the substrate phonons. Let us consider a pair of localised states $i$ and $j$. The heat current absorbed from (or released to) 
the substrate phonon bath by an electron hopping from $i$ to $j$ reads $I_{ij}^{Q}=\left(E_j-E_i\right)I^N_{ij}$, where 
$I^N_{ij}=\Gamma_{ij}-\Gamma_{ji}$ is the hopping particle current between $i$ and $j$. The local heat current associated to the 
state $i$ is given by summing over the hops from $i$ to all the states $j$:
\begin{equation}
\label{local-current}
I_{i}^{Q}=\sum_j I_{ij}^Q = \sum_{j} \left(E_j-E_i\right)I^N_{ij}.
\end{equation}
In Fig.~\ref{FIG4}, we show 2D histograms of the local heat currents $I^Q_i$ as a function of the position $x_i$ of the state $i$ 
inside the NW. We take the convention that $I^Q_{i}$ is positive (negative) when the phonons are absorbed (emitted), 
thus heating (cooling) the electrons at site $i$. We have numerically solved the random resistor network (see subsection~\ref{RRN}) 
for a temperature $k_BT=0.5t$, $W=t$ and four different values of $V_g$, corresponding to electron injection at the band center ($V_g=0$), 
below the band centre ($V_g=t$) and around the lower ($V_g=2.25t$) and upper ($V_g=-2.25t$) band edges of the NW conduction band.
At the band centre, the fluctuations of the local heat currents are symmetric around a zero average. They are larger near the NW 
boundaries and remain independent of the coordinate $x_i$ otherwise. Away from the band center, one can see that the fluctuations are no 
longer symmetric near the NW boundaries, though they become symmetric again far from the boundaries. When the electrons are injected through 
the NW in the lower energy part of the NW band, more phonons of the substrate are absorbed than emitted near the source electrode. The effect 
is reversed when the electrons are injected in the higher energy part of the band (by taking a negative gate potential $V_g$): It is now near 
the drain that the phonons are mainly absorbed. The 2D histograms corresponding to $V_g=2.25t$ and $-2.25t$ are symmetric by inversion with 
respect to $I^Q_i=0$. 

As explained in Refs.~\cite{bosisio3,bosisio4}, activated transport near the band edges of disordered NWs opens interesting 
ways for managing heat at small scales, notably for cooling hot spots in microprocessors. Taking $M=2.10^5$ NWs 
contacting two $1\mathrm{-cm}$ long electrodes, one estimates that we could transfer $0.15\,\mathrm{mW}$ from the source side towards the drain side 
by taking $\delta\mu/e \approx 1\,\mathrm{mV}$ at $T=77\,\mathrm{K}$. Again, the larger is $M$ or $\delta\mu$, the larger is the heat of the substrate 
which can be transfered from the source towards the drain by hot electrons.   

\section{Activated Multi-terminal thermoelectric transport and Ratchet effects} 
\label{section_ratchet}

\begin{figure}[h]
  \centering
  \includegraphics[keepaspectratio,width=\columnwidth]{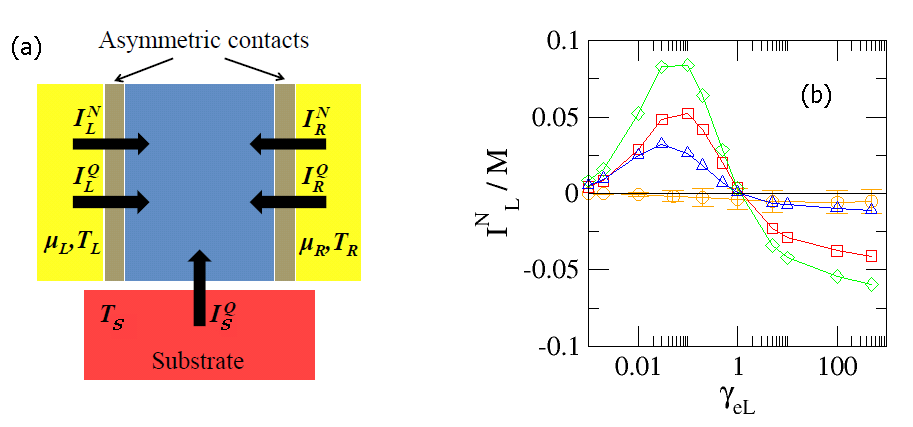}
  \caption{ ({\bf a}): Scheme of the three-terminal setup corresponding to activated thermoelectric transport for a NW deposited 
on a substrate. The NW (blue) is connected to two electronic electrodes (yellow) $L$ (the source) and $R$ (the drain) via asymmetric contacts. 
The electrodes are at equilibrium (electrochemical potentials $\mu_L$, $\mu_R$, and temperatures $T_L$, $T_R$). The insulating substrate 
(red) provides a phonon bath of temperature $T_S$. ({\bf b}): Ratchet effect powered by $\delta T_S=10^{-3}\,\epsilon$ when 
$\delta T= \delta \mu=0$. For various values of $V_g$ ($V_g/\epsilon=0$ ({\large{\color{orange}$\circ$}}), $0.5$ ({\scriptsize{\color{red}$\square$}}), 
$1$ ({\large{\color{green}$\diamond$}}) and $2.5$ ({\scriptsize{\color{blue}$\triangle$}})), the average particle currents $I^N_L/M$ 
(in unit of $10^5 \epsilon/\hbar$) of an array of $M=2.10^5$ parallel NWs is given as one varies $\gamma_{eL}$ (elastic coupling to 
the source), keeping the same value $\gamma_{eR}=\epsilon/\hbar$ for the elastic coupling to the drain. $I^N_L/M \neq 0$, unless 
$V_g=0$ (particle-hole symmetry) or $\gamma_{eL}=\gamma_{eR}$ (inversion symmetry).}
   \label{FIG5}
\end{figure}

In Secs.~\ref{section_inelastic} and~\ref{section_HotSpot}, we have discussed activated transport in a configuration where the source, drain and substrate 
were at the same equilibrium temperature $T$, the heat and particle currents between the source and the drain being induced by a small voltage bias 
$\delta \mu$ and/or temperature bias $\delta T$. More generally, a disordered NW deposited on a substrate can be viewed as the three-terminal setup 
sketched in Fig.~\ref{FIG5} ({\bf a}): Two electronic reservoirs $L$ (source) and $R$ (drain) at equilibrium with electrochemical potentials $\mu_L$, 
$\mu_R$, and temperatures $T_L$, $T_R$, while the substrate provides a third reservoir $S$ of phonons at a temperature $T_S$. Heat and particles 
can be exchanged between $L$ and $R$, but only heat with $S$. The particle currents $I^N_L$, $I^N_R$, and the heat currents $I^Q_L$, $I^Q_R$, $I^Q_S$ 
are taken positive when they enter the NW from the three reservoirs. The drain is chosen as reference ($\mu_R\equiv E_F$ and $T_R\equiv T$) 
and we set $\delta\mu = \mu_L-\mu_R,~~\delta T=T_L-T_R,~~\mathrm{and}~~\delta T_S=T_S-T_R$. In linear response the charge and heat currents 
$I^N_L$, $I^Q_L$, and the heat current $I^Q_S$ can be expressed \`a la Onsager in terms of the corresponding driving forces
\begin{equation}
\begin{pmatrix}
I^N_L \\
I^Q_L \\
I^Q_S
\end{pmatrix}
=
\begin{pmatrix}
L_{11} & L_{12} & L_{13} \\
L_{12} & L_{22} & L_{23} \\
L_{13} & L_{23} & L_{33}
\end{pmatrix}
\begin{pmatrix}
\delta\mu / T \\
\delta T / T^2 \\
\delta T_S / T^2
\end{pmatrix}\,.
\label{eq_onsager}
\end{equation}
The Casimir-Onsager relations $L_{ij}=L_{ji}$ for $i\neq j$ are valid in the absence of time-reversal symmetry breaking.
In Ref.~\cite{bosisio5}, we have discussed several possibilities offered by this setup when $\delta T_S\neq 0$, in terms of energy harvesting 
and cooling. Let us focus on the case where the NW is deposited on a hotter substrate without bias and temperature difference between 
the source and the drain ($\delta \mu= \delta T =0$ while $\delta T_S > 0$). If the particle-hole symmetry and the left-right inversion symmetry are 
broken, the heat provided by the phonons can be exploited to produce electrical work. Let us consider a model where the NW localized states 
are uniformly distributed in space and energy within a band $[-2 \epsilon, 2 \epsilon]$  with a constant DOS $\nu=1/(4 \epsilon)$ and an 
energy independent localisation length $\xi=4$. We have solved the corresponding RRN for an ensemble of $M$ parallel NWs with 
asymmetric elastic couplings to the electronics reservoirs. In average over an ensemble of $M$ NWs, particle-hole symmetry is broken if the 
Fermi potential $E_F\equiv 0$ does not coincide with the NW band centre ($V_g\neq0$), and inversion symmetry is broken by taking different 
elastic coupling constants in Eq.~\eqref{elastic-coupling} ($\gamma_{e,L}\neq \gamma_{e,R}$). Fig.~\ref{FIG5} ({\bf b}) gives the average particle 
current $I^N_L/M$ between the source and the drain as a function of $\gamma_{e,L}$ when $\gamma_{e,R}=1$ (in units of $\epsilon/\hbar$). $I^N_L/M$ was 
induced by a temperature difference $\delta T_S = 10^{-3} \epsilon$ between the substrate and a deposited array of $M=2.10^5$ parallel NWs. One can see 
that  $I^N_L \approx 0$ at the band centre ($V_g =0$) and when $\gamma_{e,L}= \gamma_{e,R}$, while $I^N_L \neq 0$ otherwise. Other ways of breaking 
inversion symmetry which give rise to even larger currents $I^N_L$ are discussed in Ref.~\cite{bosisio5}.

\section{Conclusion} 
\label{conclusion}

Our studies were restricted to the elastic tunnel regime and to the Mott VRH regime where electron-phonon interactions dominate. 
We have considered arrays of parallel purely 1D NWs where the electron states are localised and neglected the effects of 
electron-electron and electron-photon interactions. In bulk 3D amorphous germanium, silicon, carbon and vanadium oxide~\cite{Ambegaokar1971}, 
the Mott VRH behaviour was observed in a temperature range $60K <T<300 K$. This shows us that Mott VRH is relevant in a broad 
temperature domain which can reach room temperature. At lower temperatures, Efros-Shklovskii  hopping behaviour was observed in 2D 
heterostructures~\cite{Khondaker} with a universal prefactor, indicating that the effects of electron-electron interactions become more 
relevant than these of electron-phonon interactions as one decreases the temperature~\cite{fleishman}. Eventually, let us note that in bulk weakly-doped 
crystalline semiconductors, the electrons can be activated from the impurity band up to the conduction band when the temperature exceeds 
the energy gap between these two bands, putting an upper limit to VRH transport. The width of this energy gap can vary from one material to 
another. This is what can be said for bulk 3D or 2D materials. To extend these conclusions to 1D NWs is not straightforward, since the 
effect of disorder (Anderson localisation) and of electron-electron interactions (Wigner glass) are much more relevant in the 1D 
limit. Moreover, the temperature range where our predictions apply can vary from one semiconductor to another, and can depend on the width 
of the NWs. 

 In summary, we have shown that arrays of 1D MOSFETs have good thermoelectric performances, notably when the electrochemical potential is 
in the vicinity of the NW band edges. The described effects could provide an interesting method for converting waste heat into useful electrical power and for cooling hot 
spots in microprocessors. Ratchet effects open interesting perspectives when electrical transport in deposited NWs can be activated by the 
phonons of the substrate. The studied devices are based on standard nanofabrication technologies which are widely developed in 
semiconductor microelectronics. 


\end{document}